\documentstyle[twocolumn,aps]{revtex}
\begin{document}
\draft
\title{No Far--Infrared--Spectroscopic Gap in Clean and Dirty High-T$_C$
Superconductors}
\author{D. Mandrus$^{1}$\cite{dman}, Michael C. Martin$^{1}$, C. Kendziora$^{1}
$, D. Koller$^{1}$, L. Forro$^{2}$\cite{lfor}, and L. Mihaly$^{1}$}
\address{$^{1}$Department of Physics, SUNY at Stony Brook, Stony Brook,  NY
$11794$--$3800$\\
$^{2}$Department of Physics, Ecole Polytechnique Federale de Lausanne,
$1015$ Lausanne, Switzerland\\}
\date{13 October 1992}
\maketitle
\begin{abstract}
We report far infrared transmission measurements on single crystal samples
derived from Bi$_{2}$Sr$_{2}$CaCu$_{2}$O$_{8}$.  The impurity
scattering rate of the samples was varied by electron-beam irradiation, 50MeV
 $^{16}$O$^{+6}$ ion irradiation,
heat treatment in vacuum, and Y doping.  Although substantial changes in the
infrared spectra were produced, in no case was a feature observed that could
be associated with the superconducting energy gap.  These results all but
rule out ``clean limit'' explanations for the absence of the spectroscopic gap
in this material, and provide evidence that the superconductivity in
Bi$_{2}$Sr$_{2}$CaCu$_{2}$O$_{8}$ is gapless.
\end{abstract}


\narrowtext

The existence of a superconducting energy gap in the high-T$_{C}$
superconductors has been hotly debated.  A simple $s$-wave BCS model has a
complete gap of width $2\Delta $ around the Fermi energy with $2\Delta
/k_{B}T_{C}= 3.5$ for weak
coupling or higher for stronger coupling.  This energy gap is evident in the
far-infrared -- microwave range for low-T$_{C}$ superconductors
\cite{glover}.  Infrared
studies \cite{lf&glc,kamaras,schles,romero91,hughes,notgap} on high-T$_{C}$
materials reveal a feature at $\sim 8-12k_{B}T_{C}$, originally thought to be
the gap.  However there is increasing evidence to the contrary
\cite{notgap}.  In high $T_{c}$
superconductors the gap might not show up in the infrared spectrum for
several reasons. First, the scattering rate of the charge carriers may be low
relative to the superconducting energy gap $(1/\tau  << 2\Delta )$, and the
infrared spectroscopy can not distinguish
between a near perfect conductor and a true superconductor \cite{kamaras}.
Second, there seems to be a temperature
independent contribution to the oscillator strength in this freaquency range,
overlapping and possibly
masking the weak gap feature.  Third, the gap feature may be broad, either
due to anisotropy or lifetime effects.  It is also possible that the
description of the superconducting state cannot be put in a BCS framework.

We can experimentally test the first possibility by enhancing the impurity
scattering rate to produce optimum circumstances for the
observation of the spectroscopic gap.
In this Letter, we present far-infrared data on ultra-thin single crystals
of the Bi$_{2}$Sr$_{2}$CaCu$_{2}$O$_{8}$ family.  We first measured the
``pure'' samples'
far infrared transmission spectra.  The samples were then made ``dirty'' by
electron-beam irradiation, 50MeV $^{16}$O$^{+6}$ ion irradiation, or
heat treatment in vacuum to drive out oxygen.  The infrared measurements were
then repeated and compared with the ``pure'' measurements.  In another set of
experiments we looked at Y doped samples of composition
Bi$_{2}$Sr$_{2}$Ca$_{1-x}$Y$_{x}$Cu$_{2}$O$_{8}$,
with $x = 0.0 - 0.35$.  In spite of the enhanced scattering rate, the results
from all samples show no spectroscopic gap.  We present a simple analysis and
conclude that an $s$-wave BCS gap should
lead to significant features in the infrared spectrum.

The far-infrared measurements were made at beamline $U4IR$ at the National
Synchrotron Light Source, Brookhaven National Laboratory \cite{gpw}.  The
samples were mounted on the cold-finger of an $LT-3-110A$ Heli-Tran liquid
transfer refrigeration system and were maintained at various temperatures.
A Nicolet $20F$ rapid scan FTIR spectrometer with a helium cooled Si bolometer
detector recorded the spectra.  The samples used in
this study were typically $2000$\AA  thick with a diameter of $0.6mm$, and had
initial (``clean state'') transition temperatures of $T_{c}\sim 84$K.  Prior
to the infrared study the samples were characterized with electron microscopy
and dc transport measurements as described elsewhere \cite{lf&dm}.

The electron-beam irradiation was done for 20 hours in the JEOL electron
microscope at the Earth and Space Sciences Department at Stony Brook.  Although
the resistivity of the sample increased, the critical temperature was
suppressed by less than 1K.  The 50MeV $^{16}$O$^{+6}$ ion irradiation was
done at the $+30^{o}$ beam line of the Tandem Van de Graaf accelerator
operated by the Nuclear Structure Laboratory at Stony Brook \cite{noe}.
Approximately $6\times 10^{15}$ ions$/cm^{2}$ went through the sample and the
$T_{C}$ was lowered by $\sim 4K$.  The increase of the critical current,
reported earlier \cite{mcm} also demonstrated the presence of lattice damage.
The heat treatment was done with a micro-furnace while the sample was under
vacuum in the spectrometer.  The sample was heated to $\sim 650^{o}C$ to drive
out oxygen.  The critical temperature, as determined from dc resistivity
measurement, was reduced by $\sim 20K$.  The Y-doping is described in
previously published studies \cite{k,dm92}.  At the highest doping level
discussed here $(x=0.35)$ the critical temperature was about 60K.

A simple, universal indicator of the lattice damage in all samples is
the room temperature resistivity ratio, $\alpha  = \rho _{\hbox{dirty}} /
\rho _{\hbox{clean}}$.  The infrared
data can be self consistently used to calculate this quantity.  As we
established earlier \cite{lf&glc}, for low transmission samples the low
frequency limit of the transmission is $t = \rho ^{2}d^{2}(c/2\pi )^{2}$,
where $\rho $ is the dc resistivity, and $d$
is the sample thickness.   Therefore, for the samples where direct
resistivity measurements were not performed, we used $\alpha  =
\{t(300K)_{\hbox{dirty}} / t(300K)_{\hbox{pure}}\}^{1/2} = \rho _{\hbox{dirty}
} / \rho _{\hbox{pure}}$.

To verify that our samples are not inhomogeneous, a Meisner fraction
measurement would be optimal.  However
the very small and ultra-thin dimensions of our samples
make this measurement unfeasable.
Fortunately, the absolute value of the transmission at zero frequency is
another good measure of the homogeneity of our crystals.  If a sample
has a non-superconducting portion the transmission would have a non-zero
intercept.  Since we do observe in all cases that our
samples' superconducting transmissions extrapolate to zero at zero frequency,
we can assert that there are no normal state windows in the samples.

In $s$-wave BCS superconductors the frequency dependence of the
transmission coefficient of electromagnetic radiation exhibits a peak at a
frequency somewhat above (but close to) the gap frequency $\omega =2\Delta
 /\hbar $ \cite{fir}. For high-T$_C$ superconductors, this frequency is
expected to fall within the $400-800cm^{-1}$ range, depending on the gap value.

Figure \ref{fig1}(a) shows the infrared transmission spectra obtained at 13K,
100K, and 300K for the electron-beam irradiated sample.  The dotted and solid
curves were obtained before and after the irradiation, respectively.  It is
clear that for the irradiated sample the infrared transmission has increased
at all temperatures, indicative of a higher scattering rate and/or reduced
carrier density.  Looking at the 13K spectra where the sample is well below
T$_C$, we see no features representative of a gap appearing in the irradiated
sample up to $700cm^{-1}$.

The 50MeV $^{16}$O$^{+6}$ ion irradiated sample's spectra are shown in figure
\ref{fig1}(b).  Again the dotted curves display spectra taken before
irradiation and
the solid lines were measured after.  Data were obtained at 5K, 100K, and
300K.  The transmission increased significantly more for samples irradiated
this way, showing that we are doing far more damage to the sample than the
electron-beam irradiation did.  This ``dirtier'' sample again shows no evidence
of a gap in the superconducting (5K) spectra.  The 100K spectra were fit to a
simple Drude model with a mid-infrared absorption as was done previously
\cite{lf&glc}.  The fit produced the following numerical results:  Before
irradiation, $1/\tau  = 170cm^{-1}$, $\omega _{p}= 9100cm^{-1}$; after
irradiation, $1/\tau  = 230cm^{-1}$, $\omega _{p}= 6600cm^{-1}$.  The Drude
plasma frequency $(\omega _{p})$ decreased by about 40\%, corresponding to a
decrease of carrier density.  The increase of the Drude scattering rate
$(1/\tau )$ indicates that we are indeed moving away from the ``clean limit''.
If we associate the increase of the scattering rate with impurities, we obtain
$1/\tau _{0}=60cm^{-1}$ for the impurity scattering rate.

The infrared transmission of the reduced oxygen sample due to heat
treatment in vacuum is shown in Figure \ref{fig1}(c).  The transmission
increased at all temperatures (6K, 100K, and 300K) as a result of the heat
treatment.  As in the previous cases, we do not observe any feature
indicative of a superconducting energy gap.

In Fig. \ref{fig2}, we show the ratio of the transmission in the
superconducting state (taken at the lowest temperature) to the
non-superconducting spectra (taken at 100K) for all
samples in this study, including the Y doped samples
\cite{ydope}.  The investigation of the electrical transport properties of
Bi$_{2}$Sr$_{2}$Ca$_{1-x}$Y$_{x}$Cu$_{2}$O$_{8}$  indicated that the increase
of scattering rate in this approximately follows Matthiessen rule
\cite{k,dm92}.  Therefore we have reason to
believe that impurity scattering acts as a temperature independent additive
contribution to the relaxation rate.  For a sample at $x=0.38$ the resistivity
at 100K is increased by a factor of 6 relative to the pure sample, suggesting
a relaxation rate in the range of $1,200cm^{-1}$.  This number may decrease by
as much as 50\% if the changes in carrier concentration are taken into account
{\it [see figures 1 and 2 of ref \cite{dm92}]}.  Nevertheless, the normal
state relaxation rate corresponds to $1/\tau \agt 2\Delta $, and the high
relaxation rate is expected to
survive in the superconducting state.  Figure \ref{fig2} illustrates that
no peak is seen for any of the impure samples.

Earlier IR data \cite{lf&glc,kamaras,schles,romero91,hughes} saw
evidence for a gap feature at $\sim 700cm^{-1}$.  However since this feature
does not have a temperature dependance, nor does it disapear above T$_{C}$
when the T$_{C}$ is significantly
reduced, many (including us) argue that it is not the superconducting gap
\cite{notgap}.  We also argue that the gap is not hidden by the clean state
of the sample as orriginally suggested by Kamaras {\it et.al} \cite{kamaras}.
Even if the relaxation rate of the normal component drops further in the
superconducting state (as proposed by Romero {\it et. al.} \cite{romero92}),
the impurity scattering rate of our samples would be significant enough to
expose the gap feature.  In this respect our work is in contradiction to
Brunel {\it et. al.} \cite{brunel} who observed a gap feature in
the reflectivity of dirty Bi$_{2}$Sr$_{2}$CaCu$_{2}$O$_{8}$ films.

To elucidate what the lack of a peak in our data
implies, we numerically fabricated several $\sigma
_{1}(\omega )$ functions (Figure \ref{fig3}a) and calculated the resulting
transmission ratio (Figure \ref{fig3}b).  The model parameters were chosen to
represent a homogeneous, dirty
sample (using numbers generated by our fits to the $^{16}O^{+6}$ irradiated
sample).  First we simulated the optical conductivity at $100K$ as a sum of
a Drude term with $1/\tau  = 230cm^{-1}$, $\omega _{p}= 6600cm^{-1}$ and a
broad mid infrared resonance with $\Gamma  = 25000cm^{-1}$, $\Omega _{p}= 25500
cm^{-1}$ and $\omega _{0}= 3000cm^{-1}$ (Fig. \ref{fig3}a,
dashed line).  We emphasize that in the investigated frequency range the
resulting optical conductivity $\sigma _{1}(\omega )$ can not be distinguished
from that obtained by the assumption of a single carrier with frequency
dependent relaxation rate, as discussed by Schlesinger {\it et. al.} \cite
{schles}, so the particular
choice of mathematical representation does not restrict our arguments.

Next, we created a hypothetical non-superconducting low temperature
conductivity, by using  $1/\tau _{0} = 60cm^{-1}$ \cite{tau}.  To represent
superconductivity we introduced various, somewhat arbitrary cut-offs to
$\sigma _{1}(\omega )$ at low frequencies.
The missing area was calculated, a Kramers-Kr\"{o}nig transformation was
performed to provide $\sigma _{2}(\omega )$, the superconducting condensate
was represented by
the appropriate $1/\omega $ contribution, and finally the transmission in the
superconducting and non-superconducting cases were calculated and ratioed.
The first $\sigma _{1}$ function (plus signs) has a complete gap.  This type
of optical
conductivity is expected in the ``single carrier model'', where the normal
state relaxation rate is frequency dependent, there is no extra mid infrared
absorption, and the superconducting gap is complete.  We also tried a smooth
drop in $\sigma _{1}$ where no complete gap opens.  When the slope at $2\Delta
$ is large (asterisks), a pronounced peak still results.  If the transision
is very gradual (triangles), a peak is still visible, but it is broadened
significantly.  A ratio without a peak is observed when we assume a gapless
optical conductivity in the superconducting state  (squares) which smoothly
approaches the non-superconducting $\sigma _{1}(\omega )$ at higher frequency.
This gapless calculated ratio certainly best matches our data.

The general conclusions we can draw are that
gapless spectra do fit our data, and if a gap feature does exist, it must be
much smoother than predicted by the BCS theory.  This result is consistent
with a recent reflectivity study on Ni-doped YBa$_{2}$Cu$_{3}$O$_{7-\delta }$
films \cite{lemberger}.  The absence of fully developed $s$-wave gap is in
accordance with the the results of tunneling studies on the same material
\cite{dm&jh}.  The results of photoemission spectroscopy are compatable with
the absence of a $s$-wave BCS gap \cite{photo}.  The strongest argument for
a fully develloped gap came from early penetration depth studies, but recent
prececion measurements by Hardy {\it et. al.} \cite{hardy} provide evidence
for nodes in the gap function.  The anisotropy of the NMR relaxation rate
\cite{bulut}, and the nonvanishing low frequency Raman absorbtion
\cite{staufer} point in a similar direction.  Our data is another piece of a
growing amount of evidence that the low temperature density of states in the
High-$T_{C}$ materials is profoundly different from conventional
superconductors.

\acknowledgments

This work has been supported by NSF Grant \#9016456.  The NSLS was supported
by the U.S. DOE under contract \#DE$-$AC$02$--$76$CH$00016$, LF by the
Fonds National Suisse de la Recherche Scientifique $\#4030$--$032779$.
The help of R. Lefferts in $^{16}$O$^{+6}$ irradiation is appreciated.
Discussions with D. Romero are also acknowledged.

\begin{figure}
\caption{Infrared transmission of pure (dotted line) and ``dirty''
(solid line) samples.  (a) Sample before and after electron-beam irradiation.
Spectra for three temperatures are shown.  (b) Spectra for a 50MeV
$^{16}$O$^{+6}$ ion
irradiated sample for three temperatures.  The sample was irradiated with
$6\times 10^{15}$ ions$/cm^{2}$. (c) Infrared transmission for a pure and heat
treated in vacuum sample for three temperatures.}
\label{fig1}
\end{figure}

\begin{figure}
\caption{Ratios of superconducting transmission (low temperature) to
non-superconducting transmission (100K).  The curves are offset for clarity,
and the dashed line represents the transmission ratio equal to unity for each
curve.  The top panel (a) was obtained from the results presented in Figure
1; the lower panel (b) shows the results for four Bi$_{2}$Sr$_{2}$Ca$_{1-
x}$Y$_{x}$Cu$_{2}$O$_{8}$ samples
$(x$ ranging from 0 to 0.35).  The vertical scale is indicated separately on
each panel.  The quantity  $\alpha $  is a measure of sample purity, as
discussed in the text $(\alpha =1$ corresponds to pure sample).}
\label{fig2}
\end{figure}

\begin{figure}
\caption{Calculations of the superconducting to non-superconducting
transmission ratio expected for various, arbitrary cut-offs at low frequency
in $\sigma _{1}(\omega )$.  The upper panel (a) shows the non-superconducting
$\sigma _{1}(\omega )$ (dashed
line) and the different cut-offs we fabricated.  The lower panel (b) shows
the transmission ratios calculated from the $\sigma _{1}(\omega )$ of the same
symbol.}
\label{fig3}
\end{figure}


\begin {references}
\bibitem[*]{dman}Present Address:  Los Alamos National Laboratory, Los Alamos,
 New Mexico  $87545$.
\bibitem[\dagger ]{lfor}Permanent Address:  Institute of Physics of the
University, $41001$ Zagreb, Croatia.
\bibitem{glover}R.E. Glover and M. Tinkham, Phys.\ Rev.\ {\bf108} 243
(1957)
\bibitem{lf&glc}L. Forro, G.L. Carr, G.P. Williams, D. Mandrus, and L. Mihaly,
Phys.\ Rev.\ Lett.\ {\bf65} 1941 (1990)
\bibitem{kamaras}K. Kamaras {\it et. al.}, Phys.\ Rev.\ Lett.\, {\bf64} 84
(1990)
\bibitem{schles}Z. Schlesinger {\it et. al.}, Phys.\ Rev.\ Lett.\, {\bf65}
801 (1990)
\bibitem{romero91}D. Romero, G.L. Carr, D.B. Tanner, L. Forro, D. Mandrus,
L. Mihaly, G.P. Williams, Phys.\ Rev.\ B {\bf44} 2818 (1991)
\bibitem{hughes}R.A. Hughes {\it et. al.}, Phys.\ Rev.\ B {\bf40} 5162 (1990)
\bibitem{notgap}A good review of the optical data is given by G.A. Thomas in
``High Temperature Superconductivity'' Eds D.P. Tunstall, W. Barford, and P.
Osborne, (Adam Higler, 1991) p. 169;  also D.B. Tanner {\it et. al.}, in
``High Temperature Superconductivity'', Eds. J. Ashkenazi, S. Barnes, F. Zuo,
G. Vezzoli, B. Klein, (Plenum Press, 1991), p. 159
\bibitem{romero92}D. Romero {\it et. al.}, Phys.\ Rev.\ Lett.\ {\bf68} 1590
(1992);D. Mandrus {\it et. al.}, Phys.\ Rev.\ B {\bf46} 8632 (1992)
\bibitem{brunel}L.C. Brunel {\it et. al.}, Phys.\ Rev.\ Lett.\ {\bf66} 1346
(1991)
\bibitem{fir}In ``Far-Infrared Properties of Solids'', Eds. S.S. Mitra and
S. Nudelman, (Plenum, 1970), p. 223
\bibitem{lf&dm}L. Forro, D. Mandrus, B. Kenszei, and L. Mihaly, J.\ Appl.\
Phys.\ {\bf68} 4876 (1990)
\bibitem{noe}John W. No\'{e} , Rev.\ Sci.\ Instrum.\, {\bf57} 757 (1986)
\bibitem{mcm}Michael Martin, C. Kendziora, L. Mihaly, and R. Lefferts, Phys.\
Rev.\ B {\bf46} 5760 (1992)
\bibitem{k}C. Kendziora {\it et. al.}, Phys.\ Rev.\ B {\bf45} 13025 (1992)
\bibitem{dm92}D. Mandrus, L. Forro, C. Kendziora, and L. Mihaly, Phys.\ Rev.\
B {\bf45} 12640 (1992)
\bibitem{gpw}G.P. Williams, Nucl.\ Instrum.\ Methods, {\bf A291} 8 (1990)
\bibitem{ydope}For the Y doped samples the direct comparison of ``before''
and ``after'' spectra is less informative, since each Y-doped specimen was
prepared from a separate doped batch and had a slightly different thickness.
Taking ratios reduces or eliminates the non-intrinsic sample-to-sample
variations due to the differences inthe sample thickness.
\bibitem{tau}This choice corresponds to
residual impurity scattering determined earlier for $^{16}$O$^{+6}$ irradiated
samples.  Assuming that $1/\tau $ is independent of frequency,
this represents a ``worst case'' scenario
for obtaining a transmission peak, in the sense that a higher relaxation
rates would produce larger peak.
\bibitem{lemberger}M.J. Sumner, J.-T. Kim, T.R. Lemberger, to be published in
Phys. Rev. B.
\bibitem{dm&jh}D. Mandrus, J. Hartge, L. Forro, C. Kendziora and L. Mihaly,
to be published in Europhysics Letters; D. Mandrus, L. Forro, D. Koller, and
L. Mihaly, Nature {\bf351} 460 (1991)
\bibitem{photo}Z.-X. Shen {\it et. al.}, Phys. Rev. Lett. {\bf 70} 1553
(1993); Y. Hwu {\it et. al}, Phys. Rev. Lett. {\bf 67} 2573 (1991);
D.S. Dessau {\it et. al.}, Phys. Rev. Lett. {\bf 66} 2160 (1991)
\bibitem{hardy}W.N. Hardy, D.A. Bonn, D.C. Morgan, Ruixing liang, and Kuan
Zhang, preprint.
\bibitem{bulut}N. Bulut and D.J. Scalapino, Phys. Rev. Lett. {\bf 68} 706
(1992)
\bibitem{staufer}T. Staufer {\it et. al.}, Phys. Rev. Lett {\bf 68} 1069
(1992); F. Slakey {\it et. al.}, Phys. Rev. B {\bf 41} 2109 (1990)
\end{references}
\end{document}